\documentclass[aps]{revtex4}%
\usepackage{amsfonts}
\usepackage{amsmath}
\usepackage{amssymb}
\usepackage{graphicx}%
\begin{document}
\title[Relaxation time distribution]{Determination of the relaxation time distribution from the dielectric losses}
\author{S.A. Ktitorov}
\affiliation{A.F. Ioffe Physico-Technical Institute of the Russian Academy of Sciences,
Polytechnicheskaja str. 26, St. Petersburg 194021, Russia}

\begin{abstract}
A problem of determining of the relaxation time distribution in irregular
relaxing systems is considered from the point of view of the
Kotelnikov-Shannon theorem of the telecommunication theory.

\end{abstract}
\maketitle









The dielectric relaxation in irregular systems like polymers, impure
ferroelectrics etc, and magnetic relaxation in impure magnetics including spin
glasses \cite{spinglass} are usually described with a help of a continuous
relaxation time distribution function (RTDF). It is desirable to have, along
with theoretical models for such distributions, some regular technique for
determining of them using the measured experimentally dielectric
$\epsilon(\omega)$ or magnetic $\mu\left(  \omega\right)  $ spectrum. We will
discuss for definiteness the former case. The exact formula relating the
dielectric losses and RTDF was derived in \cite{kirk} many years ago .
However, this formula can not be used directly in practice (see the discussion
below). Numerical analysis of the proper integral equation was done recently
\cite{schafer}, \cite{pelster}. Our aim here is to develop an approach giving
a practical procedure for determining of RTDF.

We introduce RTDF $g(\ln(\tau/\tau_{0}))$ with a help of the relation
\cite{distr}:%

\begin{equation}
\chi(\mu)=\int_{-\infty}^{\infty}\chi_{D}(\omega\tau)g(\ln(\tau/\tau_{0}%
))d\ln(\tau/\tau_{0} \label{distr}%
\end{equation}
with%

\begin{equation}
\chi_{D}(\omega\tau)=\frac{\epsilon_{D}(\omega\tau)-\epsilon_{\infty}%
}{\epsilon_{0}-\epsilon_{\infty}}=\frac{1}{1-i\omega\tau}, \label{debye}%
\end{equation}
where $\epsilon_{D}(\omega\tau)$ is the Debye single-relaxator dielectric
permittivity, $\epsilon_{0}$ and $\epsilon_{\infty}$ are respectively zero-
and high-frequency limits of the permittivity, $\tau_{0}$ is an arbitrary time
scale. Basing on rather intuitive grounds, the distribution function is often
estimated assuming $g(\ln(\tau/\tau_{0})=\operatorname{Im}\chi(\ln(\omega
\tau_{0})$. Our aim here is to give a regular procedure for calculation of the
distribution function $g(\ln(\tau/\tau_{0})$ basing on a measured dielectric
spectrum $\epsilon(\omega)$.

Let us introduce the logarithmic variables:%

\begin{align}
z  & =\ln(\tau/\tau_{0});\omega\tau_{0}=\Omega;\tau/\tau_{0}=\exp
(z),\Omega=\exp(-\mu),\label{log}\\
\omega\tau & =\Omega\exp(z)=\exp(z-\mu),y=-\ln(\omega\tau_{0}),\nonumber
\end{align}
Then equation (1) can be rewritten in the form%

\begin{equation}
\chi(y)=+\int_{-\infty}^{\infty}dx\frac{g(x)}{1-i\exp(x-y)}.
\label{difference}%
\end{equation}
Now we discuss some possible approaches to the problem. The integral equation
(\ref{difference}) can be solved exactly. This equation belongs to the class
of equations that can be solved using the Fourier transform. Taking the
imaginary part we have%

\begin{equation}
\chi^{\prime\prime}(\omega)=\int_{-\infty}^{\infty}d\ln(\tau/\tau_{0}%
)\frac{g(\ln(\tau/\tau_{0})\omega\tau}{1+(\omega\tau)^{2}}.\label{integral}%
\end{equation}
Introducing the logarithmic variables  we can write%

\begin{equation}
\chi^{\prime\prime}(\mu)=\int_{-\infty}^{\infty}dz\frac{g(z)\Omega\exp
(z)}{1+\Omega^{2}\exp(2z)}=\int_{-\infty}^{\infty}dz\frac{g(z)\exp(z-\mu
)}{1+\exp(2(z-\mu))}=\frac{1}{2}\int_{-\infty}^{\infty}dz\frac{g(z)}%
{\cosh(z-\mu)}. \label{diffinteq}%
\end{equation}
Let us write (\ref{diffinteq}) in the standard form%

\begin{equation}
\chi^{\prime\prime}(\mu)=\int_{-\infty}^{\infty}dzK(\mu-z)g(z).
\label{standard}%
\end{equation}
This integral equation can be solved using the Fourier transform:%

\begin{equation}
\chi_{y}^{\prime\prime}=\int_{-\infty}^{\infty}dz\exp(iyz)\chi^{\prime\prime
}(z), \label{direct1}%
\end{equation}

\begin{equation}
g_{y}=\int_{-\infty}^{\infty}dz\exp(iyz)g(z), \label{direct2}%
\end{equation}

\begin{equation}
K(z-\mu)=1/[2\cosh(z-\mu)]. \label{kernel}%
\end{equation}
Performing the Fourier transform we obtain an algebraic equation relating the
Fourier transformants:%

\begin{equation}
\chi_{y}^{\prime\prime}=K_{y}g_{y}. \label{algebr}%
\end{equation}
Using the known integral \cite{grad}%

\begin{equation}
\int_{0}^{\infty}dx\frac{\cos xy}{\cosh\alpha x}=\frac{\pi}{2\alpha\cosh
(\frac{\pi y}{2\alpha})}, \label{int}%
\end{equation}
we can write%

\begin{equation}
K_{y}=\pi/[2\cosh(\pi y/2)]. \label{fourierkern}%
\end{equation}
Thus the Fourier transform of the solution reads%

\[
g_{y}=2\pi^{-1}\chi_{y}^{\prime\prime}\cosh(\pi y/2).
\]
Taking the inverse Fourier transformation%

\begin{equation}
g(z)=\int_{-\infty}^{\infty}\frac{dy}{2\pi}\exp(-iyz)g_{y}, \label{inverse}%
\end{equation}
we can, in principle, calculate RTDF basing on the measured dielectric losses
function $\chi^{\prime\prime}(\mu).$ These integrals must be computed
numerically if the susceptibility is presented as a point numerical function.
The trouble with this way is the extremely poor convergence of the integral
with respect to $\mu$. However, there is another possible way: if we can carry
out an analytic continuation of the dielectric losses function we can
integrate generically. We assume the integrals to be uniformly convergent
within the domain of integration so that one can change the order of
integration (an exact analysis must be carried out on the base of the theory
of distributions \cite{schwartz}):%

\[
g(z)=\frac{1}{2\pi^{2}}\int_{-\infty}^{\infty}dy\int_{-\infty}^{\infty}%
d\mu\exp[(\mu-z)yi]\chi^{\prime\prime}(\mu)\left[  \exp(\pi y/2)+\exp(-\pi
y/2)\right]  =
\]

\[
=\frac{1}{\pi}\int_{-\infty}^{\infty}d\mu\chi^{\prime\prime}(\mu)\left[
\delta(z-\mu+i\pi/2)+\delta(z-\mu-i\pi/2)\right]  =
\]

\begin{equation}
=\frac{1}{\pi}\left[  \chi^{\prime\prime}(z+i\pi/2)+\chi^{\prime\prime}%
(z-i\pi/2)\right]  \label{solv}%
\end{equation}
This solution was derived by Kirkwood \cite{kirk} in a different way.

Now we discuss possible ways for practical use of this formula. Assuming
differentiability of $\chi^{\prime\prime}(z)$ up to at least the second
derivative and its slowliness and expanding it in powers of $i\pi/2,$ we can
give the approximate form of the solution:%

\begin{equation}
g(z)\approx\frac{2}{\pi}\left[  \chi^{\prime\prime}(z)-\frac{\pi^{2}}{8}%
\frac{d^{2}\chi^{\prime\prime}(z)}{dz^{2}}+\ldots\right]  . \label{appr}%
\end{equation}
\qquad\qquad This approximate formula is asymptotically exact for very
slowly-varying dielectric functions. Formula \ref{solv} can be basically used
in a general case, but ambiguity of the analytic continuation makes it
problematic. There are two main reason for it. Firstly, results of the
analytic continuation essentially depend on analytic properties of the
function, but they cannot be established exactly basing on a function
determined on the finite segment of the real axis. Secondly, measuring of the
dielectric losses function yields really a discrete set of points that
presents an additional ambiguity factor. This means that analytic properties
must be determined independently. A similar problem exists in the
communication theory. Discretization of a signal and impossibility to transfer
a signal with unlimited spectrum through the channel make it necessary to
apply the approach developed in the communication theory by Kotelnikov and
Shannon \cite{kotel} (basically similar approach was really known earlier in
the mathematical theory of approximation). Following to the spirit of this
approach, we assume that $\chi^{\prime\prime}(z)$ is an entire function of
finite degree \cite{kotel}:%

\begin{equation}
\lim_{r\rightarrow\infty}\frac{\ln\max_{\left\vert z\right\vert =r}\left\vert
\chi^{\prime\prime}(z)\right\vert }{r^{2}}=\alpha<+\infty\label{finite}%
\end{equation}
But according to the Paley-Wiener theorem \cite{difference}, such a function
has a bounded spectrum. What does it mean from the physical point of view in
our case? The spectrum of $\chi^{\prime\prime}(z)$ characterizes the time
evolution of the polarization; therefore, boundness of the spectrum means that
a reaction of the polarization on the field pulse decays for some finite time.
Surely, this decay time is theoretically infinite in glasses, but it is
essentially restricted by noise and other factors in a real experiment. That
is why we have right to assume $\chi^{\prime\prime}(z)$ to be an entire
function of finite degree.

Let $\chi^{\prime\prime}(z)$ be a function with the spectrum $g_{y}$ given by
\ref{direct2} lying in the band $-\beta<y<\beta.$ Then the following
interpolating formula holds for $\chi^{\prime\prime}(z)$ at $\alpha>\beta$:%

\begin{equation}
\chi^{\prime\prime}(z)=\sum_{k=-\infty}^{\infty}\chi^{\prime\prime}(k\frac
{\pi}{\alpha})\frac{\sin\alpha\left(  z-k\frac{\pi}{\alpha}\right)  }%
{\alpha\left(  z-k\frac{\pi}{\alpha}\right)  }. \label{interpol}%
\end{equation}
This series converges uniformly to the entire analytic function and admits an
analytic continuation. Substituting (\ref{solv}) into (\ref{interpol}) we come
to the final formula:%

\[
g(z)=\frac{1}{\pi}\sum_{k=-\infty}^{\infty}\chi^{\prime\prime}(k\frac{\pi
}{\alpha})\left[  \frac{\sin\alpha\left(  z+i\frac{\pi}{2}-k\frac{\pi}{\alpha
}\right)  }{\alpha\left(  z-k\frac{\pi}{\alpha}\right)  }+\frac{\sin
\alpha\left(  z-i\frac{\pi}{2}-k\frac{\pi}{\alpha}\right)  }{\alpha\left(
z-k\frac{\pi}{\alpha}\right)  }\right]  =
\]

\begin{equation}
\frac{\cosh\frac{\alpha\pi}{2}}{\pi}\sum_{k=-\infty}^{\infty}\chi
^{\prime\prime}(k\frac{\pi}{\alpha})\frac{\sin\alpha\left(  z-k\frac{\pi
}{\alpha}\right)  }{\alpha\left(  z-k\frac{\pi}{\alpha}\right)  }
\label{solv1}%
\end{equation}

Thus, we have shown that for the case of the finite-time response functions,
the "naive" solution $g(z)\propto\chi^{\prime\prime}(z)$ is exact within the
class of the entire analytic functions. Notwithstanding the fact that a real
measurement gives only finite-time response because of the noise effect, it
would be interesting to generalize this result to the case functions, which
have an infinite spectrum with power-law decreasing. We hope to return to this
problem in the future.

\end{document}